\documentclass[showpacs,amsmath,amssymb,twocolumn,prl,floatfix]{revtex4}
\usepackage[dvips]{graphicx}
\input{epsf}

\begin{document}

\title{Loschmidt cooling by time reversal of atomic matter waves}

\author{J. Martin, B. Georgeot and D. L. Shepelyansky}

\affiliation{Laboratoire de Physique Th\'eorique,
 Universit\'e de Toulouse III, CNRS, 31062 Toulouse, France}

\date{October 25, 2007; Revised: November 28, 2007}
%\date{\today}

\begin{abstract}
We propose an experimental scheme which allows to realize
approximate time reversal of matter waves for ultracold atoms in
the regime of quantum chaos.  We show that a significant fraction of
the atoms return back to their original state, being at the same
time cooled down by several orders of magnitude.  We give a theoretical
description of this effect supported by extensive numerical
simulations.
The proposed scheme can be implemented with existing experimental setups.
\end{abstract}
%05.45.Mt Quantum chaos; semiclassical methods
%03.75.Lm Tunneling, Josephson effect, Bose-Einstein condensates in periodic
%    potentials, solitons, vortices, and topological excitations
%32.80.Lg Mechanical effects of light on atoms, molecules, and ions
%32.80.Pj Optical cooling of atoms; trapping
%32.80.Qk Coherent control of atomic interactions with photons
%03.75.-b Matter waves (for atom interferometry techniques, see 39.20.+q - in
%    atomic and molecular physics)

\maketitle

%%%%%%%%%%%%%%%%%%%%%%%%%%%%%%%%%%%%%%%%%%%%%%%%%%%%%%%%%%%%%%%%%%%%%

The statistical theory of gases developed by Boltzmann leads to
macroscopic irreversibility and entropy growth even if dynamical equations of
motion are time reversible.  This contradiction was pointed out by Loschmidt
and is now known as the {\em Loschmidt paradox} \cite{loschmidt}.  The reply
of Boltzmann relied on the technical difficulty of velocity reversal
for material particles \cite{boltzmann}:  a story tells that
he simply said ``then go and do it''.  The modern resolution
of this famous dispute came with the development of the theory of
dynamical chaos \cite{sinai,chir1,lieberman}.  Indeed, for chaotic
dynamics
small perturbations grow exponentially with time, making the motion
practically irreversible.  This explanation is valid for classical dynamics,
while the case of quantum dynamics requires special consideration.
Indeed, in the quantum case the exponential growth takes place only
during the rather short Ehrenfest time \cite{chirikov}, and the
quantum evolution remains stable and reversible in presence of small
perturbations \cite{dima83}.  Quantum reversibility
in presence of various perturbations has been actively studied
in recent years and is now described through the {\em Loschmidt echo}
(see \cite{prosen} and Refs.~therein).
This quantity measures the effect of perturbations
and is characterized by the fidelity
$f(t_r)=|\langle \psi_p (2t_r) |\psi (0)\rangle|^2$, where
$|\psi_p \rangle $ is
the time reversed wavefunction
in presence of perturbations,
$|\psi \rangle$ is the unperturbed one
and $t_r$ is the moment of time reversal.
Experimental implementations of
time reversibility for quantum dynamics or propagating waves
have been realized with spin systems (spin echo technique) \cite{hahn},
acoustic \cite{finkac} and electromagnetic \cite{finkem} waves,
resulting in various technological applications.
Surprisingly enough, the reversibility signal becomes stronger and
more robust in the case of chaotic ray dynamics \cite{finkac}.
However, despite the significant experimental progress made recently
in the control of quantum systems, the time reversal of matter waves
has not been performed so far.

Here we present a concrete experimental proposal of
an effective time reversal of atomic matter waves.  The proposal relies
on the kicked rotator model, which is a cornerstone model of quantum
chaos \cite{chirikov,dima83,fishman}.  This model
has been built up experimentally with cold atoms in kicked optical
lattices \cite{raizen,christensen,darcy,garreau}.  Recent progress allowed
to implement this model with ultracold atoms
and Bose-Einstein Condensates (BEC)
\cite{phillips,summy,steinberg,sadgrove}, with
high-precision subrecoil definition of the momentum of the atoms,
allowing for example to observe \cite{phillips,steinberg}
high order quantum resonances \cite{izrailev}.  We show that these
experimental techniques allow to perform time reversal
for a significant part of the atoms.  Surprisingly, this fraction
of the atoms becomes cooled down by several orders of magnitude during the
process.  We call this new cooling mechanism {\em Loschmidt cooling}
since it is directly related to the time reversibility.

The quantum kicked rotator corresponds to the quantization of
the Chirikov standard map \cite{chir1,lieberman}:
\begin{equation}
\label{map}
\bar{p} = p + k \sin{ x} \; , \;\;
\bar{x} = x + T \bar{p}
\end{equation}
where $x$ is the position and $p$ the momentum of an atom,
and bars denote the variables after
one  map iteration. Here $x$ is a continuous
variable in the interval $] -\infty,+\infty [$.
The physical process behind corresponds to
rapid change of momentum created by a kick of optical lattice
followed by a free propagation of the atoms between periodic kicks
of period $T$.
The classical dynamics depends only on the single parameter
$K=kT$, and undergoes a transition from integrability to chaos
when $K$ is increased. Global chaos sets in
 for $K >  K_c = 0.9716...$.  The dynamics of (\ref{map}) is
time reversible, e.g.~by inverting all velocities
at the middle of the free motion between two kicks.

The quantum evolution over one period
is described by
a unitary operator $\hat{U}$ acting on a wavefunction:
\begin{eqnarray}
\label{qmap}
\bar{\psi} = \hat{U} \psi = e^{-iT\hat{p}^2/2} e^{-ik\cos{\hat{x}}} \psi
\end{eqnarray}
where the momentum $p$ is measured in recoil units, and $T$ plays the
role of an effective Planck constant.
The momentum operator $\hat{p}=-i \partial / \partial x $
 has eigenvalues $p=n+\beta$ where
$n$ is an integer and
$\beta$ is the quasimomentum for a wave propagating in the $x$ direction.
The particle energy is $E=E_r p^2/2$ where $E_r$ is the recoil energy.
We consider noninteracting atoms.
It is convenient to express the time $t$ in number
of map iterations. In experiments, values of time up to
$t=150$ have been achieved \cite{garreau}.
Also, a very narrow initial momentum distribution
down to rms
$\sigma_{\beta} \approx  0.002$ can be reached with BEC
\cite{phillips}.  Values as high as $k \approx 4$ have been
realized experimentally with $T$ varying between $1$ and $4\pi$
\cite{garreau}.

To perform the time reversal, we first write $T$ as $T=4\pi+\epsilon$.
After $t_r$ iterations of (\ref{qmap}), we interchange the order
of kick and free propagation, change $T$ to
$T'=4\pi-\epsilon$ and $k$ to $-k$, and let the system evolve
during another $t_r$ iterations.  Such a modification of $T$ can be
easily realized experimentally (see e.g.~\cite{phillips}).  The sign
of $k$ can be inverted by changing the sign of the detuning
between the laser and the atomic transition frequencies or through a shift
of the optical lattice by half a wavelength.
Then for $t>t_r$ the map (\ref{qmap}) becomes
 $\hat{U}_r=e^{ik\cos{\hat{x}}}e^{-iT'\hat{p}^2/2}$ where the second operator
acts on momentum eigenstates $|n+\beta\rangle$ as
$e^{-iT'(n+\beta)^2/2}=e^{i T(n+\beta)^2/2} \; e^{-8i\pi \beta (n+\beta/2)}$.
Thus $\hat{U}_r= \hat{U}^{\dagger} \; e^{-8i\pi \beta(\hat{n}+\beta/2)}$
and the components with $\beta=0$ (integer values
of $p$) are exactly reversed, while for other small $\beta$ values
the time reversal works only approximately.  This determines the
meaning of {\em approximate} time reversal.
Another
way of inverting $k$ is to use in the first and last
propagation steps for $t>t_r$ the value $T''=2\pi-\epsilon$ instead of $T'$.
Indeed for $\beta=0$ we have
$e^{i\pi \hat{n}^2}e^{-ik\cos{\hat{x}}}e^{i \pi \hat{n}^2}=
e^{ik\cos{\hat{x}}}$, and thus inversion of $k$ works again exactly for $\beta=0$
and approximately for small $\beta$ values.  In the following, we use
the first method of $k$ inversion, but we checked that the second
method gives similar results.

\begin{figure}
\begin{center}
\includegraphics[width=.95\linewidth]{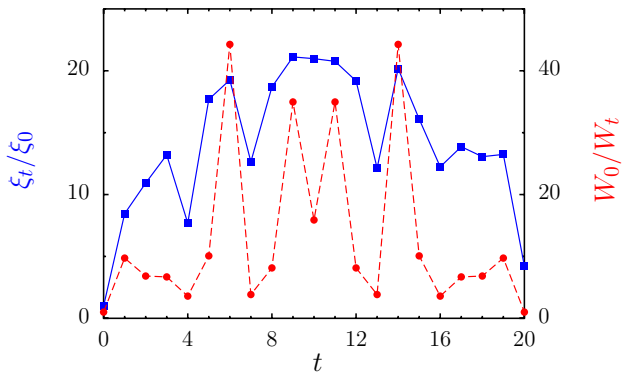}
\end{center}
\vglue -0.60cm
\caption{(Color online) Dependence on time $t$
of the rescaled IPR $\xi_{t}/\xi_0$
(solid curve with blue/black squares) and of the
rescaled inverse probability at zero momentum
$W_0/W_{t}$ (dashed curve with red/gray dots).
Initial state is a Gaussian wave packet
with rms $\sigma_{\beta}=0.002$ and
$k_B T_0/E_r=\sigma_{\beta}^2/2=2\times 10^{-6}$
(see Fig.~\ref{fig2}).
Here $k=4.5$, $T=4 \pi+\epsilon$ for
$0\leq t \leq t_r$ and $T'=4 \pi-\epsilon$ for $t_r < t \leq 2t_r $
with $\epsilon=2$ and $t_r=10$.}
\label{fig1}
\end{figure}

To characterize the wave packet dynamics,
we use the Inverse Participation Ratio (IPR) defined
by $\xi_t= \sum_p W_p (t)/\sum_p W_p^2(t)$ where
$W_p(t)=|\langle p |\psi (t)\rangle |^2$
is the probability in momentum space at time $t$.
Another useful quantity is the probability at zero
momentum $W_t=W_0(t)$.
For numerical simulations we used up to $N=2^{23}$ discretized
momentum states with $\Delta p=1/40000$.
Both quantities $\xi_t$ and $W_t$ are rescaled by their values
at $t=0$ that makes them independent of $\Delta p$. Their dependence
on time is shown in Fig.~\ref{fig1} which clearly demonstrates
time reversal of both quantities.  The return for $\xi_t$ is not
perfect due to the contribution of nonzero quasimomentum components, whereas
the curve for $W_t$ is perfectly symmetric.  We note that $\xi_t$ shows
a diffusive growth for $t<4$ followed by a saturation due to
dynamical localization \cite{chirikov,dima83,fishman,raizen} for $t \leq t_r$.

\begin{figure}
\begin{center}
\includegraphics[width=.95\linewidth]{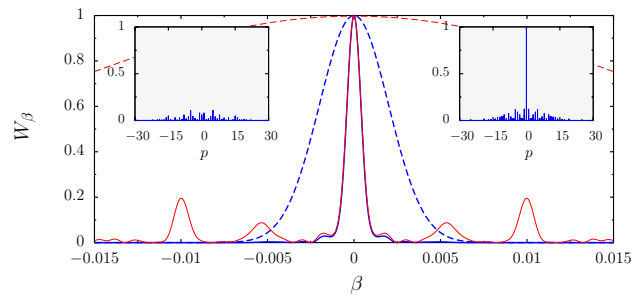}
\end{center}
\vglue -0.40cm
\caption{(Color online)
Initial states probability (dashed curves) and final
return probability (solid curves) shown as a function
of quasimomentum $\beta$.  Here
$k_BT_0/E_r =2 \times 10^{-4}$ (red/gray curves) and
$k_BT_0/E_r =2 \times 10^{-6}$
(blue/black curves), the latter corresponding to data of Fig.~\ref{fig1}
and being similar to experimental conditions \cite{phillips}.
For return probabilities, the central peaks coincide
for the two values of $T_0$.
Insets: probability
$W_p$ at $t=t_r=10$ (left) and final
probability $W_p$ at $t=t_f=2t_r$ (right) on a larger scale with $p=n+\beta$
for $k_BT_0/E_r =2 \times 10^{-6}$.
All distributions are scaled by the value of
the initial probability at zero momentum ($\beta=0$).
Parameters are as in Fig.~\ref{fig1}.}
\label{fig2}
\end{figure}

The momentum probability distributions at initial $t=0$ and final $t=t_f=2t_r$
times are shown in Fig.~\ref{fig2}.  The striking feature is that
the final distribution in $-0.5 <p< 0.5$
is much
more narrow than the initial one and has the same
maximal value since the probability at $\beta=0$ returns
exactly.  The shape of the reversed peak is independent of $T_0$.
This can be interpreted as a cooling of the atoms
which remained in this momentum range, that defines
the {\em Loschmidt cooling} mechanism.  The narrowing of the central peak
means that in compensation a significant fraction of atoms has obtained
higher momentum values $p$ as is clearly seen in the right inset of
Fig.~\ref{fig2}.  But even if the full distribution is rather broad
the reversed peak is clearly dominant.  On the contrary,
the left inset showing the distribution at $t=t_r$ displays
homogeneous chaotic distribution of momentum components.
Thus it is the time reversal which produces the peak at the
origin and performs effective cooling.  It is natural to define
the size of the return Loschmidt peak by its half width $\beta_L$
with $W_{\beta_L}(2t_r) =W_{\beta=0}(2t_r)/2$.
The fraction of returned atoms is
$P_\beta=\sum_{-2\beta_L}^{2\beta_L} W_{\beta}(2t_r)$ and their temperature
is $T_f=\sum_{-2\beta_L}^{2\beta_L} \beta^2 W_{\beta}(2t_r)/(2P_{\beta})$.
Similarly to the case of chaotic acoustic cavities \cite{finkac},
quantum chaos makes the time reversed peak more visible.
From an experimental viewpoint, the atoms outside the reversed
peak can be eliminated by an escape process while those inside
can be kept by switching on a suitable trap potential or
attraction between atoms (e.g. Feschbach resonance).
Such a procedure is similar to
the process of evaporative cooling. However in our case the
effective evaporation takes place very rapidly due to dynamical chaos.

\begin{figure}
\begin{center}
\includegraphics[width=.95\linewidth]{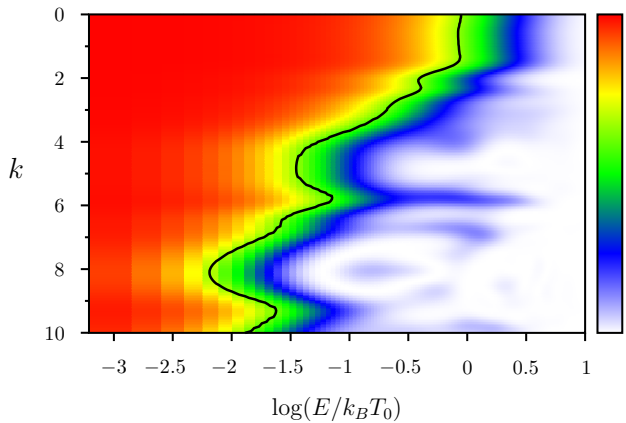}
\end{center}
\vglue -0.40cm
\caption{(Color online) Density plot of the return probability distribution
$W_{p} (2t_r)$ as a function of the rescaled atom energy $E/k_BT_0$ and of
the kick strength $k$, where $E=E_r p^2/2$ and $k_BT_0/E_r =2 \times 10^{-6}$;
here $\epsilon=2$ and $t_r=10$.
Colors denote density from white (minimal) at the right
to red/gray (maximal) at the left.
Black curve shows the final temperature $T_f$ of the return Loschmidt peak
($E\rightarrow k_B T_f=\langle E \rangle$) as a function
of $k$. Logarithm is decimal.}
\label{fig3}
\end{figure}

The variation of the final return probability distribution in energy
$E=E_r p^2/2$
with $k$ is shown in Fig.~\ref{fig3}. When $k$ increases the
distribution in energy becomes more and more narrow, so that
Loschmidt cooling becomes more efficient.  This is related
to the fact that the dynamics becomes more chaotic as $k$ increases.
The temperature $T_f$ drops
by almost two orders of magnitude, showing significant
oscillations with $k$.

\begin{figure}
\begin{center}
\includegraphics[width=.95\linewidth]{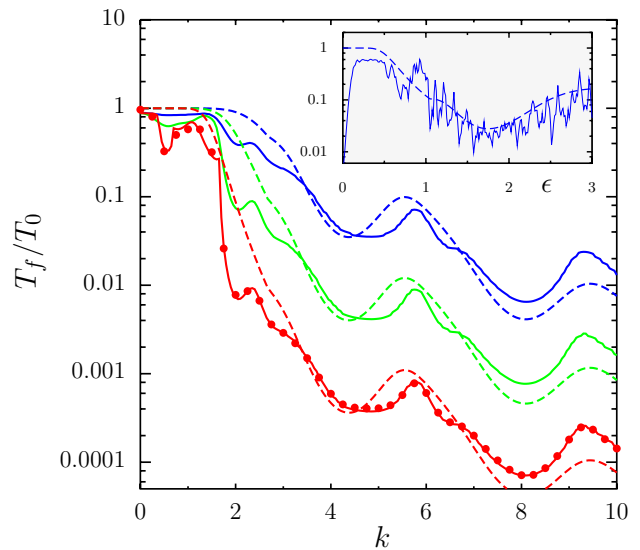}
\end{center}
\vglue -0.40cm
\caption{(Color online) Ratio of final to initial temperatures
(solid curves) as a function of $k$, from top
to bottom for $k_BT_0/E_r =2 \times 10^{-6}$
(blue/black), $k_BT_0/E_r =1.8 \times 10^{-5}$ (green/light gray),
and $k_BT_0/E_r =2 \times 10^{-4}$ (red/gray), with
$\epsilon=2$ and $t_r=10$.  Dashed curves show the theory
(\ref{temp}). Full circles show $P_{\beta}^2$ for
$k_BT_0/E_r =2 \times 10^{-4}$, confirming
theory (\ref{pbeta}).  Inset : same
ratio as a function of $\epsilon$
 for $k=4.5$; solid curve shows numerical data,
dashed curve is the theory (\ref{temp}).}
\label{fig4}
\end{figure}

The decrease of $T_f/T_0$ with $k$ is shown in more details in
Fig.~\ref{fig4}.   It is related to the increase of the localization
length $l$ of quasienergy eigenstates with $k$.
Indeed, it is known that $l= D_q/2$ where
$D_q =k^2 g(K_q)/2$ is the quantum diffusion rate,
$K_q=2k \sin (T/2)$ being the quantum chaos parameter \cite{chirikov,dima87}.
The function $g(K_q)$ takes into account the effects of quantum
correlations and is given \cite{dima87} by $g(K_q)=0$ for $K_q < K_c$,
$g(K_q)\approx 0.42 (K_q-K_c)^3 /K_q^2$ for $K_c \leq K_q < 4.5$
and
$g(K_q)= 1-2J_2(K_q)-2J_1^2(K_q)+2J_2^2(K_q)+2J_3^2(K_q)$ for $K_q \geq 4.5$
where $J_m$ are Bessel functions.
Due to this localization there is always a residual probability
$W_{\beta}\sim 1/ l\sigma_{\beta}$
in the interval $-0.5<p<0.5$, even in absence of time reversal.
However this residual probability is much smaller than the maximum
of the reversed peak $W_0 \sim 1/\sigma_{\beta}$.
The width of this peak can be estimated as follows:
$\beta > 0$ in $\hat{U}_r$ acts as a small perturbation
of the exactly reversible operator, whose eigenstates have $M \sim l$
components.  This perturbation gives after time $t_r$ an accumulated
 quantum phase shift in the wavefunction
$\Delta \phi=8\pi\beta n t_r \approx 8\pi\beta M t_r$. Thus
only the atoms with $\beta \leq \beta_L \sim 1/(8 \pi M t_r)$
return to their initial state, and their
fraction is $P_{\beta} \sim  \beta_L/\sigma_{\beta}$.
 The Loschmidt temperature of these
atoms is $k_B T_L = E_r \beta_L^2/2 \approx E_r/ (128 \pi^2 C D_q^2 t_r^2+1)$
where $C$ is a numerical constant
which according to our data is $C \approx 0.4$.
Thus the ratio $T_f/T_0$ is:
\begin{equation}
\label{temp}
T_f/T_0=T_L/(T_0+T_L) \; ,
k_B T_L\approx  E_r/ (500 D_q^2 t_r^2+1)
\end{equation}
The formula (\ref{temp}) interpolates between the weakly perturbative
regime with $l \ll 1$ and the strong chaos regime with $l \gg 1$.
This theory assumes that $k_B T_0 \ll E_r$ and that
the localization time scale $t^* \approx D_q$
is shorter than $t_r$ which is approximately satisfied for most
$k$ values in Fig.~\ref{fig4}.
The theory (\ref{temp}) satisfactorily describes the global behavior
of $T_f/T_0$ as can be seen in Fig.~\ref{fig4}. Small scale oscillations
should be attributed to mesoscopic fluctuations.  These fluctuations
are stronger when $\epsilon$ is varied (see inset of Fig.~\ref{fig4}),
that should be attributed to high order quantum resonances
\cite{izrailev} at rational
values of $T/4\pi$ \cite{note}.  We checked that the cooling remains
robust even in presence of $1\%$ fluctuations of
$\epsilon$ during map iterations.
For the case $t_r \ll t^*$ one
should use $M \approx \sqrt{D_q t_r}$ instead of $M =D_q$ since the
diffusion takes place during the whole time interval $t_r$.  In this case,
$k_B T_L\approx E_r/ (128 \pi^2 C D_q t_r^3+1)$.
\begin{figure}
\begin{center}
\includegraphics[width=.95\linewidth]{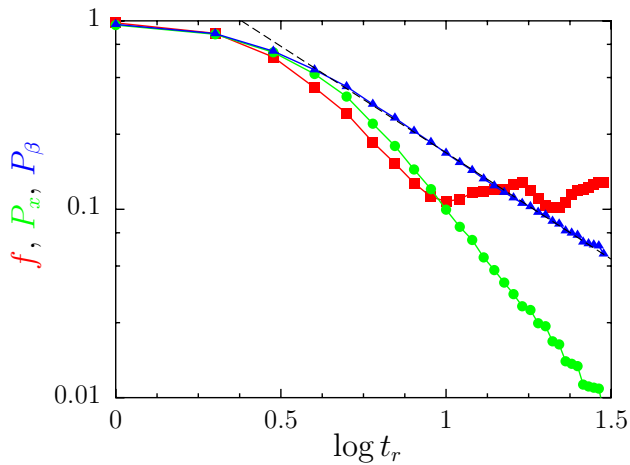}
\end{center}
\vglue -0.40cm
\caption{(Color online) Fidelity $f$ (squares),
fraction $P_\beta$ of atoms in the Loschmidt peak (triangles)
and fraction $P_x$ of these atoms in the coordinate
space cell (circles) (see text) as a function of time reversal
moment $t_r$.
Straight dashed curve shows the fit $\log P_{\beta}= -1.13\log t_r +0.43$.
Parameters are as in Fig.~\ref{fig1}.
Logarithms are decimal.}
\label{fig5}
\end{figure}

It is important to evaluate the fraction $P_{\beta}$
of returning atoms.
The estimates given above lead
to $P_{\beta} \sim \beta_L/\sigma_{\beta}
\propto 1/(\sigma_{\beta}D_q t_r)$ and thus:
\begin{equation}
\label{pbeta}
T_f/T_0 \approx P_{\beta}^2 ,
\end{equation}
that is well confirmed by the data in Fig.~\ref{fig4}.
The formula
(\ref{pbeta}) is written for dimension $d=1$.  For higher
dimensions it generalizes to $T_f/T_0 \approx P_{\beta}^{2/d}$.
The data
displayed in Fig.~\ref{fig5} show that $P_{\beta}$ drops approximately
as $1/t_r$, in agreement with the estimate above.
For example, the cooling by a factor of 100 is reached for $10\%$ of
atoms.
We also checked that a significant fraction of the atoms returns
not only in momentum space but also in coordinate space.
To this end we define the fraction of atoms $P_x$ which are
in the return Loschmidt peak and in the coordinate
space interval $[-2/\sigma_{\beta},2/\sigma_{\beta}]$.
The data show that $P_x$ is close to the value of $P_{\beta}$ for moderate
values of $t_r \le 10$,
so that the Loschmidt cooled atoms remain close both in momentum and
coordinate space and thus can be efficiently captured by a trap potential
or a Feschbach resonance.
We note that at large $t_r$ values $P_{\beta}$ can become smaller
than the fidelity
$f=|\langle \psi(0)| \psi(2t_r) \rangle|^2$, which characterizes
proximity of the whole wavefunctions and not only the reversed peak.

In conclusion, we have presented a concrete experimental proposal
of time reversal of matter waves of ultracold atoms in the regime
of quantum chaos.  If the atoms were classical particles, they would
never return back due to exponential instability of dynamical chaos.
But the quantum dynamics is stable and thus a large fraction of the atoms
returns back even if the time reversal is not perfect.  This fraction
of the atoms exhibits Loschmidt cooling which can decrease their
temperature by several orders of magnitude.
The reversed peak is very sensitive to variations of $\beta$ and other
parameters breaking time reversal symmetry, and therefore this setup
can be used as a sensitive Loschmidt interferometer to explore such a
symmetry breaking (e.g.~a gravitational field component
along the optical lattice gives a shift in
$\beta$ which affects the Loschmidt peak).
The parameters considered
here are well accessible to nowadays experimental setups (see e.g.
\cite{raizen,christensen,darcy,garreau,phillips,summy,steinberg,sadgrove}).
The realization of our proposal will shed a new light on the
long-standing Boltzmann-Loschmidt dispute on time reversibility.

We thank CalMiP for access to their supercomputers and
the French ANR (project INFOSYSQQ) and the EC project EuroSQIP for support.

\vglue -0.50cm


\begin{thebibliography}{99}
\bibitem{note} At $\epsilon=0$ there is also a significant
                cooling related to ballistic propagation at the main
                quantum resonance; however, this effect takes place
                only in a very narrow $\epsilon$ range making it rather
                sensitive to perturbations.
\end{thebibliography}
\end{document}